\documentclass[journal,10pt]{IEEEtran}
\usepackage{ifpdf}
\usepackage{cite}
\usepackage{amsmath}
\usepackage{array}
\usepackage{booktabs}
\usepackage{stfloats}
\usepackage{cite}
\usepackage{graphicx}
\usepackage{amsmath}
\usepackage{epsfig}
\usepackage{epstopdf}
\usepackage{algorithm}
\usepackage{algorithmicx}
\usepackage{algpseudocode}
\usepackage{blkarray}
\usepackage[table,xcdraw]{xcolor}
\usepackage{float}
\usepackage{amsfonts}
\usepackage{subcaption}
\usepackage{mathrsfs}
\usepackage{amssymb}
\usepackage{graphicx}
\usepackage[utf8x]{inputenc}
\usepackage{multirow,bigdelim}
\usepackage{dsfont}
\usepackage{ragged2e}
\usepackage{nicematrix}
\usepackage{mathtools}
\usepackage{hhline}
\usepackage{balance}
\usepackage{tabularray}
\usepackage{soul}

\graphicspath{{figures/}}

\newtheorem{lemma}{Lemma}

\begin{document}
	
\title{Access-based Lightweight Physical Layer Authentication for the Internet of Things Devices}
	
\author{Saud Khan,~\IEEEmembership{Student Member,~IEEE}, Chandra Thapa,~\IEEEmembership{Member,~IEEE}, Salman Durrani,~\IEEEmembership{Senior Member,~IEEE}, and Seyit Camtepe,~\IEEEmembership{Senior Member,~IEEE
\thanks{Saud Khan and Salman Durrani are with the School of Engineering, The Australian National University, Canberra, 2601, Australia (Email: \{saud.khan, salman.durrani\}@anu.edu.au)
}
\thanks{Saud Khan, Chandra Thapa and Seyit Camtepe are with Data61, Commonwealth Scientific and Industrial Research Organization (CSIRO), Sydney, 2122, Australia (Email: \{chandra.thapa, seyit.camtepe\}@data61.csiro.au)
}
\thanks{This research was undertaken with the assistance of resources and services from the National Computational Infrastructure (NCI), which is supported by the Australian Government.
}
\thanks{This article will be presented in part at the IEEE GLOBECOM Workshop 2023 \cite{Khan2312}.}
}
}

%\markboth{IEEE Internet of Things Journal}
{}%
	
\maketitle

\IEEEpeerreviewmaketitle

\begin{abstract}
Physical-layer authentication is a popular alternative to the conventional key-based authentication for internet of things (IoT) devices due to their limited computational capacity and battery power. However, this approach has limitations due to poor robustness under channel fluctuations, reconciliation overhead, and no clear safeguard distance to ensure the secrecy of the generated authentication keys. In this regard, we propose a novel, secure, and lightweight continuous authentication scheme for IoT device authentication. Our scheme utilizes the inherent properties of the IoT devices’ transmission model as its source for seed generation and device authentication. Specifically, our proposed scheme provides continuous authentication by checking the access time slots and spreading sequences of the IoT devices instead of repeatedly generating and verifying shared keys. Due to this, access to a coherent key is not required in our proposed scheme, resulting in the concealment of the seed information from attackers. Our proposed authentication scheme for IoT devices demonstrates improved performance compared to the benchmark schemes relying on physical channels. Our empirical results find a near threefold decrease in the misdetection rate of illegitimate devices and close to zero false alarm rate in various system settings with varied numbers of active devices up to 200 and signal-to-noise ratio from 0 dB to 25 dB. Our proposed authentication scheme also has a lower computational complexity of at least half the computational cost of the benchmark schemes based on support vector machine and binary hypothesis testing in our studies. This further corroborates the practicality of our scheme for IoT deployments.

\end{abstract}	

\section{Introduction}
Internet of Things (IoT) devices are increasingly indispensable to modern society, industry, and governments \cite{shahab2020grant}. These devices are expected to increase from 14 billion in 2022 to 27 billion in 2025 \cite{iot_analytics_2022}. Moreover, these devices will form an integral part of future networks, including 6G~\cite{6G_iot}. However, the security of these widely used IoT devices is an increasingly essential issue~\cite{9279294}.

IoT devices usually connect to a network through an access point (AP). The conventional approach to establish secure communications between IoT devices and AP is to generate a shared secret key by exploiting the reciprocity of the random fading channel \cite{hershey1995unconventional, patwari2009high}. Herein, the IoT devices measure highly correlated wireless channel characteristics (\emph{e.g.}, channel impulse responses, or received signal strengths) and use them as shared random sources to generate a shared key. However, the low-cost and often resource-constrained IoT devices cannot facilitate physical-channel probing for a shared key generation due to the limited resources. Instead, these IoT devices rely on intermittent transmissions, which makes them highly susceptible to adversarial attacks~\cite{xu2016security}.

Besides, IoT devices perform sporadic transmission to save energy. Considering sporadic transmission and the massive number of IoT devices in the future network, the non-orthogonal multiple access (NOMA) transmission protocols, which overlap multiple IoT devices over a single radio resource block, are envisioned as a potential solution~\cite{shahab2020grant, 10178037}. Though the sporadic transmission in NOMA enables IoT devices to transmit for extended periods, it also negates the suitability of conventional shared key generation schemes for authentication due to their inherent complexity and reliance on shared key updates. As a result, there can be potential adversaries with abundant opportunities for certain attacks, such as spoofing attacks and eavesdropping~\cite{stellios2018survey, hassan2019current}. 

Existing methods, such as upper-layer security protocols, suffer from high computational overhead~\cite{lin2017survey}. Conversely, lightweight options are available, but these often rely on physical channel attributes~\cite{10012482, 9851464} and are unreliable in the presence of variations and noises. Furthermore, channel probing is challenging given the resource limitation of the devices. This underscores the need for fast (enabled by continuous authentication mechanism), reliable (no reliance on physical channel attributes), and lightweight authentication mechanisms for IoT devices.

\subsection{Contributions}
In this paper, we propose a novel lightweight and continuous authentication scheme for resource-constrained IoT devices by identifying the pre-arranged access time slots and spreading pools of each IoT device, which provides high uncertainties for the spoofers and supplies seamless protections for legitimate communications. In our proposed scheme, the access time slots are pre-agreed between a pair of IoT devices and the AP, which are difficult for the adversaries to predict and do not require additional hardware for implementation \cite{9378574}. The access time slots are generated using the spreading pools available at the AP and IoT devices. The access time slots for every IoT device are generated independently at the AP and the IoT devices, thereby obeying the grant-free NOMA protocol for a practical massive IoT deployment. If the access time slot and spreading pool of an IoT device are different from the access time slot and spreading pool at the AP, it will be identified as an illegitimate device by the AP.  \textit{To our best knowledge, this is the first work to authenticate multiple resource-constrained IoT devices utilizing grant-free NOMA protocol by utilizing their spreading pools and pre-arranged access time slots as the source for authentication.} The main contributions of this work are summarised as follows.

\begin{itemize}
    \item \textbf{Authentication scheme}: We propose a lightweight authentication scheme comprised of four processes: access time slots generation, spreading pool construction, seed generation, and authentication decision. The scheme provides continuous authentication by checking the access time slots and spreading pools of the IoT devices instead of generating and verifying shared keys.
    
    \item \textbf{Reduced overhead and latency}: The spreading sequences, utilized by the IoT devices as part of the grant-free NOMA transmission protocol, are used as the seed source for access time slot generation and IoT device authentication. Thus, our proposed scheme does not need seed verification and reconciliation processes, which incur massive overhead and latency.
    
    \item \textbf{Improved authentication performance}: Our results in the misdetection rate of illegitimate devices indicate a nearly threefold improvement, false alarm rate indicates state-of-the-art, and spreading sequence collision rate indicates superior performance in different settings while boasting a lower complexity compared to the benchmark schemes. Furthermore, our proposed scheme does not rely on the physical channel reciprocity assumption, which makes it a suitable authentication scheme for resource-constraint IoT devices.  
\end{itemize}

\textit{Paper organization :} The rest of this paper is organized as follows. In Section II, we review the related studies of authentication schemes for IoTs. In Section III, we present the system model and the authentication problem. In Section IV, we describe the proposed authentication scheme and provide a detailed description of the different phases of device authentication. In Section V, we derive the performance analysis of the proposed authentication scheme. In Section VI, we present the simulation results to verify the performance gain of the proposed technique. Finally, Section VII concludes the paper.

\textit{Notations:} Lower and upper case boldface letters are used for vectors and matrices, respectively. The transpose of a vector $\mathbf{a}$ is $\mathbf{a}^T$.  The norm is denoted by $\left|\left| \cdot \right|\right|$. $\mathbb{C}^{x \times y}$ denotes the complex valued space of size $x \times y$ respectively. $\odot$, $\oslash$, and $(\cdot)^\dagger$ denote the Hadamard product, the Hadamard division, and the Moore-Penrose matrix inversion, respectively. Table~\ref{tab:dimen} summarizes the important symbols used in this work, including the dimensions of vectors and matrices.

\section{Related Works}
Considering the adversaries, upper-layer security protocols have been increasingly studied in the literature \cite{lin2017survey, weber2010internet}. However, they are not well suited for resource-constrained IoT devices due to their massive computational overhead and excessive latency. In this regard, low-complexity authentication schemes are desirable for resource-constrained IoT devices, complementing the overall network entropy by introducing additional measures for IoT device authentication in the lower layers \cite{wang2016physical, paul2008physical}. 

\begin{table}
\centering
\caption{Important symbols used in this work.}
\label{tab:dimen}
\begin{tabular}{|c|l|c|} 
\hline
Variable                  & Description                                                                                                           & Dimension     \\ 
\hline
$K$                       & Total number of IoT devices                                                                                           & $1 \times 1$           \\ 
\hline
$N$                       & Total subcarriers                                                                                                     & $1 \times 1$           \\ 
\hline
$S$                       & Active number of IoT devices                                                                                          & $1 \times 1$           \\ 
\hline
$J$                       & Number of time slots                                                                                                  & $1 \times 1$           \\ 
\hline
$\mathbf{c}$              & Spreading sequence                                                                                                    & $N \times 1$  \\ 
\hline
$\mathbf{h}$              & Channel                                                                                                               & $N \times 1$  \\ 
\hline
$\mathbf{x}$              & Transmit signal                                                                                                       & $K \times 1$  \\ 
\hline
$\mathbf{w}$              & Gaussian noise                                                                                                        & $N \times 1$  \\ 
\hline
$\mathbf{y}$              & Received signal                                                                                                       & $N \times 1$  \\ 
\hline
$\mathbf{G}$              & \begin{tabular}[c]{@{}l@{}}Synthesis of channel vector and \\spreading sequences\end{tabular}                         & $N \times K$  \\ 
\hline
$\mathbf{H}$              & Channel matrix                                                                                                        & $N \times K$  \\ 
\hline
$\mathbf{C}$              & Codebook matrix                                                                                                       & $N \times K$  \\ 
\hline
$\mathbf{X}$              & Transmit signal (continuous time slots)                                                                               & $K \times J$  \\ 
\hline
$\bar{\mathbf{G}}$        & \begin{tabular}[c]{@{}l@{}}Synthesis of channel vector and \\spreading sequences (continuous time slots)\end{tabular} & $N \times K $  \\ 
\hline
$\mathbf{W}$              & Gaussian noise (continuous time slots)                                                                                & $N \times J$  \\ 
\hline
$\mathbf{Y}$              & Received signal (continuous time slots)                                                                               & $N \times J$  \\ 
\hline
$\bar{\mathbf{\Gamma}}$ & Authenticated devices' indicator                                                                                       & $K \times J$  \\ 
\hline
$\tilde{\mathbf{X}}$      & Authenticated devices' data                                                                                            & $K \times J$  \\
\hline
\end{tabular}
\end{table}

Physical layer security schemes based on keyless authentication \cite{hou2014physical, zhang2020physical, xie2021physical} can provide lightweight security to the resource-constrained IoT devices by exploiting the inherent physical-channel attributes and/or device-specific features of IoT devices. By doing so, the overall network entropy can be improved while reducing IoT devices' computational cost and energy consumption. The authors in \cite{hou2014physical} introduced scheduling policies to utilize the physical channel characteristics for device authentication. The authors in \cite{zhang2020physical} utilized the channel and phase noise of the physical channel between a transceiver pair utilizing multiple antennas for hypothesis testing and device authentication. Similarly, the authors in \cite{xie2021physical} utilized the correlation of multiple channel impulse responses (CIR) from the physical channel for authentication. Recently, machine learning has also been applied to combine with physical layer authentication schemes to improve the robustness under channel fluctuations \cite{10012482, 9851464, 9446939}. However, the reliance of these techniques on the physical channel for feature extraction results in unreliable authentication performance due to variations and noises present in complex dynamic environments.   

In a different approach, to achieve continuous authentication, the authors in \cite{fang2019learning} used an authentication mechanism to create a learning-based kernel model that utilizes multi-attributes from the physical channel for device authentication. Then, the authors in \cite{9378574} utilized the multi-attribute design of the physical channel and support vector machine (SVM) to utilize pseudo-random binary access time slots for device authentication. However, due to the time-varying nature of the physical channel, especially in complex dynamic environments, and the low-cost components utilized by the IoT devices, the variations and noise cause unreliable seed acquisition. Additionally, since these works are based on the assumption of physical channel reciprocity, they will incur a high seed mismatch rate due to the half-duplex nature of the resource-constrained IoT devices; this results in multi-staged parity bits for seed reconciliation, which is against the deployment spirit of resource-constrained IoT devices. Moreover, since the IoT devices are resource-constrained, the physical channel probing process cannot be carried out due to the inherent sporadic communication nature of the IoT devices. 

In order to overcome the aforementioned challenges, we propose an access-based framework that paves the way for lightweight and continuous authentication tailored for resource-constrained IoT devices.

\begin{figure*}[t]
\centering
\includegraphics[scale=0.57]{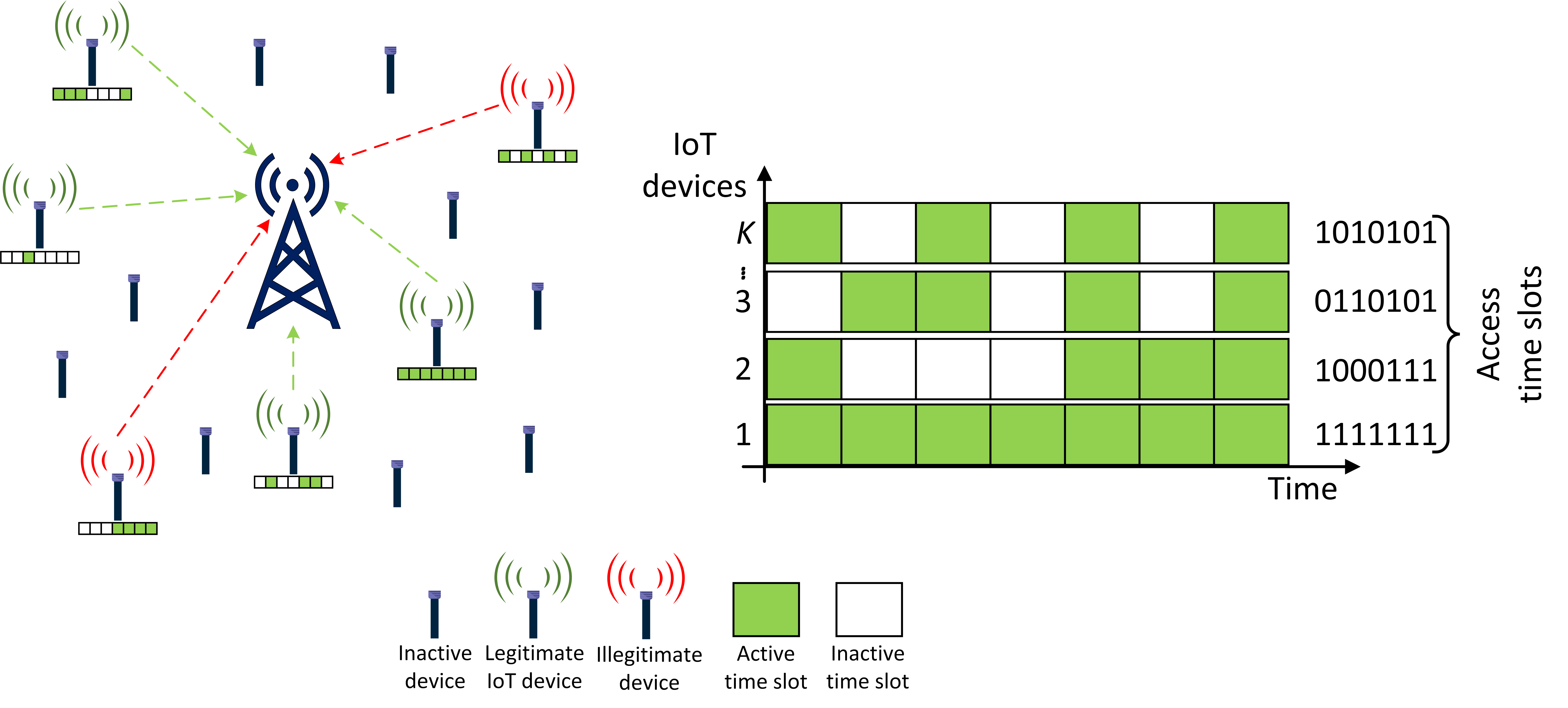} 
\caption{Illustration of our system model. The transmission between the IoT devices and the AP is carried out by following the pre-agreed access time slots.}
\label{SystemMod}
\end{figure*}

\section{System Model}
In this paper, we consider the scenario where IoT devices wake up sporadically and transmit their data to the AP in a grant-free manner, as depicted in Fig.~\ref{SystemMod}. 
Thus, we consider a spreading-based uplink grant-free NOMA system comprising of an AP and $K$ IoT devices with limited computing capabilities. The AP has relatively powerful computing capabilities and is at a fixed location. The AP and IoT devices are assumed to be equipped with a single antenna, and their clocks are synchronized\footnote{Practically, clock synchronization can be achieved via methods described in \cite{8008773, 8125129, 6774858} to achieve energy-efficient communications for IoT devices. However, this is outside the scope of this work.}. We assume that upper-layer security mechanisms are utilized initially to establish system parameters between the AP and IoT devices \cite{9446939}. During transmission, a subset of the $K$ IoT devices sporadically and randomly become active when they have data to transmit. We consider an overloaded system where the number of resource blocks $N$ is less than the number of IoT devices in a cell, \emph{i.e.}, $N < K$.

\subsection{Threat Model}
In the system model, as depicted in Fig.~\ref{SystemMod}, we assume that illegitimate devices can be present anywhere in a cell, including in close proximity to legitimate IoT devices, and therefore, their physical channels can be correlated. As a result, the AP can receive transmissions from both legitimate IoT and illegitimate devices, where the illegitimate devices attempt to access the network by conducting spoofing attacks, such as man-in-the-middle attacks and replay attacks. With this in mind, apart from the codebook matrix\footnote{Generally, the AP can refresh the codebook matrix in a cell to enhance communication using different methods \cite{koyluoglu2011interference, perazzone2018cryptographic}. However, this is a separate research topic and is, therefore, outside the scope of this work.}, we assume that the illegitimate devices utilize the same system parameters and upper-layer signaling as the legitimate IoT devices, as detailed in Table~\ref{tab:dimen}. We further assume that the illegitimate devices can remain active at all times and can scan the network to learn the transmission pattern of legitimate IoT devices. Thus, illegitimate devices can be resourceful and more computationally capable than legitimate IoT devices.       

\subsection{Signal Model}
Considering an arbitrary symbol interval, an IoT device randomly wakes up and transmits its complex modulated signal towards the AP, which are independent random variables drawn from a standard symmetric discrete constellation set. After modulation, the transmitted symbol $x_k$ from the $k$-th IoT device is spread onto a spreading sequence $\mathbf{c}_k$ of length $N$. The received signal $y$ on the $n$-th subcarrier at the AP is given as 
\begin{equation} \label{eqmain1}
        y_n  =	\sum_{k=1}^{K} h_{nk}  c_{nk} x_k   +   w_n,\\
\end{equation}
where $h_{nk}$ refer to the $n$-th subcarrier of the $k$-th IoT device's channel vector $\mathbf{h}_k = \left[  h_{1k}, h_{2k}, \hdots, h_{Nk} \right]^T \in \mathbb{C}^{N \times 1}$, $c_{nk}$ refer to the $n$-th component of the spreading sequence $\mathbf{c}_k = \left[  c_{1k}, c_{2k}, \hdots, c_{Nk} \right]^T \in \mathbb{C}^{N \times 1}$, and $w_n$ is the Gaussian noise on the $n$-th subcarrier with zero mean and variance $\sigma^2$. By combining the received signals overall $N$ subcarriers, the received signal vector $\mathbf{y} = [y_1, y_2, \hdots, y_N ]^T \in \mathbb{C}^{N \times 1}$ is given as
\begin{equation} \label{eqmain2}
        \mathbf{y}  =	\mathbf{G x + w},\\
\end{equation}
where $\mathbf{x} = [x_1, x_2, \hdots, x_K]^T \in \mathbb{C}^{K \times 1}$ is the transmitted signal vector for all $K$ devices and $\mathbf{w} = \left[  w_{1}, w_{2}, \hdots, w_{N} \right]^T \in \mathbb{C}^{N \times 1}$ is the noise vector. $\mathbf{G} \in \mathbb{C}^{N \times K}$ is the synthesis of the channel vectors and spreading sequences, given as
\begin{equation}\label{CodeBookwa}
    \mathbf{G} = \mathbf{H} \odot \mathbf{C},
\end{equation}
where $\mathbf{H} = [\mathbf{h}_1, \mathbf{h}_2, \hdots, \mathbf{h}_K] \in \mathbb{C}^{N \times K}$ is the channel matrix, $\mathbf{C} = [\mathbf{c}_1, \mathbf{c}_2, \hdots, \mathbf{c}_K] \in \mathbb{C}^{N \times K}$ is the codebook matrix, and $\odot$ is the Hadamard product, \emph{i.e.}, $g_{nk} = h_{nk} c_{nk}$.

\subsection{Transmission Model}
Different works \cite{7462187, kim2020deep, zou2021joint} have assumed that the active IoT devices remain unchanged in an entire frame. However, in practical grant-free systems, the IoT devices access or leave the system randomly \cite{shahab2020grant}. Moreover, once active, due to the size of their data payload, some IoT devices transmit their data in consecutive time slots. From this, it concurs that the nature of data transmission by IoT devices is generically random and not deterministic. Therefore, we consider a scenario where the IoT devices become active or inactive in different time slots, which is a more practical scenario in 6G IoT applications with sporadic communications. Motivated by this, we can extend the signal model in (\ref{eqmain2}) from a single time slot transmission model to a continuous time-slots transmission model.

The transmitted signals $\mathbf{X} = \left[ \mathbf{x}^{[1]}, \mathbf{x}^{[2]}, \hdots, \mathbf{x}^{[J]}   \right] \in \mathbb{C}^{K \times J} $ are recovered from the received signals $\mathbf{Y} = \left[ \mathbf{y}^{[1]}, \mathbf{y}^{[2]}, \hdots, \mathbf{y}^{[J]}   \right] \in \mathbb{C}^{N \times J}$ in $J$ continuous time slots, based on the LTE-Advanced standard protocol \cite{ETUR2021evolved}. Thus, the continuous time-slots transmission model for the $j$-th time slot is given as
\begin{equation} \label{ContiSys}
    \mathbf{y}^{[j]}  =	\mathbf{G}^{[j]} \mathbf{x}^{[j]} + \mathbf{w}^{[j]}, \quad j = 1,2, \hdots, J,
\end{equation}
where $\mathbf{G}^{[j]} \in \mathbb{C}^{N \times K}$ is the synthesis of the channel vectors and spreading sequences in the $j$-th time slot and $\mathbf{w}^{[j]}$ is the equivalent Gaussian noise vector in the $j$-th time slot.  

\subsection{Problem Statement}
The sporadic nature of the IoT devices allows the illegitimate devices to impersonate the legitimate IoT devices to spoof the AP and gain access to the core network. Assuming that an IoT device transmits to the AP in the $j$-th time slot, the objective at the AP is to authenticate the device if the message originated from a legitimate IoT device. In order to achieve this, the AP and the legitimate IoT devices can agree on specific transceiver features or characteristics, which can be used to distinguish legitimate IoT devices from illegitimate devices. Let $\mathbf{\Gamma}^{[j]}$ represent the authenticated devices indicator in the $j$-th time slot; then, the authentication problem is given as
\begin{equation} \label{ProbStat}
    \mathbf{\Gamma}^{[j]} = 
    \begin{cases}
        1 & \text{if } \mathcal{H}_0\\
        0 & \text{if } \mathcal{H}_1
    \end{cases}
    ,
\end{equation}
where $\mathcal{H}_0$ and $\mathcal{H}_1$ represents the received signal $\mathbf{y}^{[j]}$ in the $j$-th time slot, originated from a legitimate IoT device and an illegitimate device, respectively, and act as the hypothesis for IoT device authentication. The conventional schemes \cite{liu2013two, liu2016physical, zhang2020physical} rely on quantization-based thresholds in (\ref{ProbStat}) for decision making. However, the authentication performance significantly declines due to the quantization errors introduced by the algorithms. Additionally, it is challenging to obtain optimal values for the detection thresholds to maintain continuous authentication when a large number of IoT devices are involved since exhaustive search methods are utilized to obtain these values.  

Another downside to these conventional schemes is that they rely on the physical channel for seed acquisition, verification, reconciliation, and IoT device authentication \cite{wang2016physical, paul2008physical, 9279294}. However, reliance on the physical channel for device authentication does not explicitly apply to resource-constrained IoT devices. The reasons for this are as follows. 
\begin{itemize}
    \item A transceiver pair cannot probe the physical channel simultaneously for seed acquisition due to the half-duplex nature of the radio. The resource-constrained IoT devices are assumed to probe the physical channel for seed acquisition and authentication in the conventional physical-channel-based schemes. This is impractical since the resource-constrained IoT devices cannot probe the physical channel due to their limited battery; therefore, the conventional schemes result in excessive battery loss and time lag due to the radio distance turnaround time.
    \item The reconciliation overhead due to imperfect physical-channel reciprocity increases with the increased key length for seed generation. This means that to achieve a higher authentication rate (by increasing key length), the parity bit information to correct errors is also increased. This is against the spirit of authentication mechanisms for resource-constrained IoT devices, where channel training/probing of IoT devices should be minimized due to their limited resources.
    \item A transceiver pair separated by a greater than half wavelength distance does not guarantee independent physical channels for seed acquisition \cite{edman2011passive}. This means that there is no clear safeguard distance to ensure the secrecy of the generated key, as typically assumed in the physical-channel-based seed acquisition techniques \cite{zeng2015physical}.
\end{itemize}
From this discussion, we can conclude that (i) conventional physical-channel-based authentication techniques exhibit these intrinsic limitations, which limits their effectiveness in situations where a transceiver pair experiences spoofing attacks, and (ii) the conventional physical-channel-based seed acquisition techniques are not practical for resource-constrained IoT devices. Therefore, access to a coherent source for identical and lightweight seed generation is crucial for continuous authentication between the AP and resource-constrained IoT devices.

\begin{figure*}[t]
\centering
\includegraphics[scale=0.55]{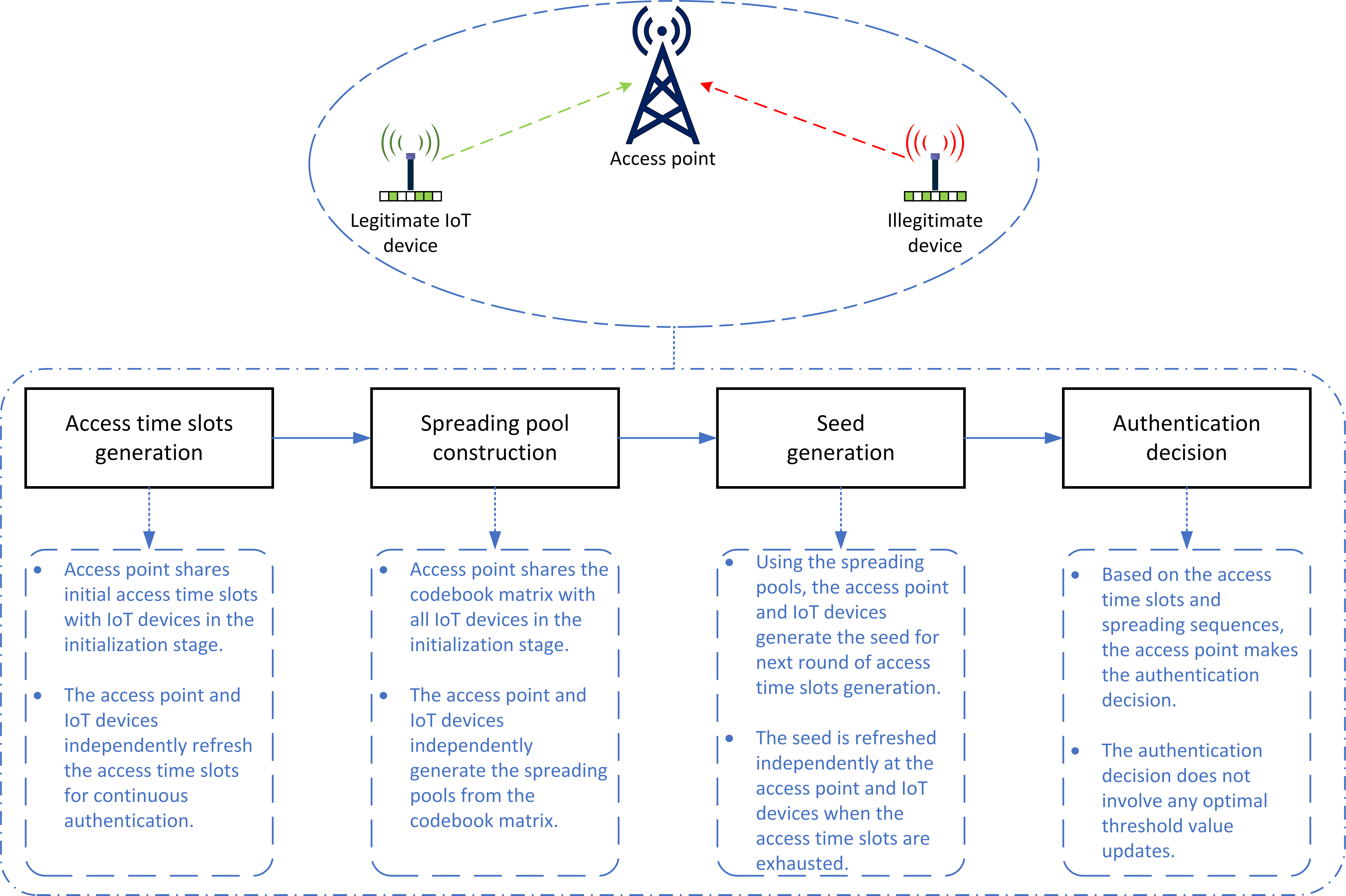} 
\caption{Proposed authentication scheme comprises four processes: access time slots generation, spreading pool construction, seed generation, and authentication decision.}
\label{PRBS}
\end{figure*}

\section{Proposed Authentication Scheme}
With the sporadic nature of transmission of IoT devices in mind, the objective at the AP is to authenticate the legitimate IoT devices from the received signal $\mathbf{y}^{[j]}$ in the $j$-th time slot. Therefore, to achieve authentication, the generated seeds must adhere to the policies as follows \cite{9378574}: 1) a transceiver pair must generate an identical seed stemming from an identical feature for authentication at the AP; 2) seeds should be undisclosed to any other devices, making the generated feature unpredictable by illegitimate devices; and 3) seeds should be proactively refreshed to maintain continuous authentication while preserving uncertainty for illegitimate devices. To meet these requirements, we use the transmission nature of the grant-free NOMA in (\ref{ContiSys}) as the seed source instead of relying on physical-channel attributes for seed acquisition. Then, we use the seed to generate the access time slots for IoT device authentication.

\begin{figure*}[t]
\centering
\includegraphics[scale=0.5]{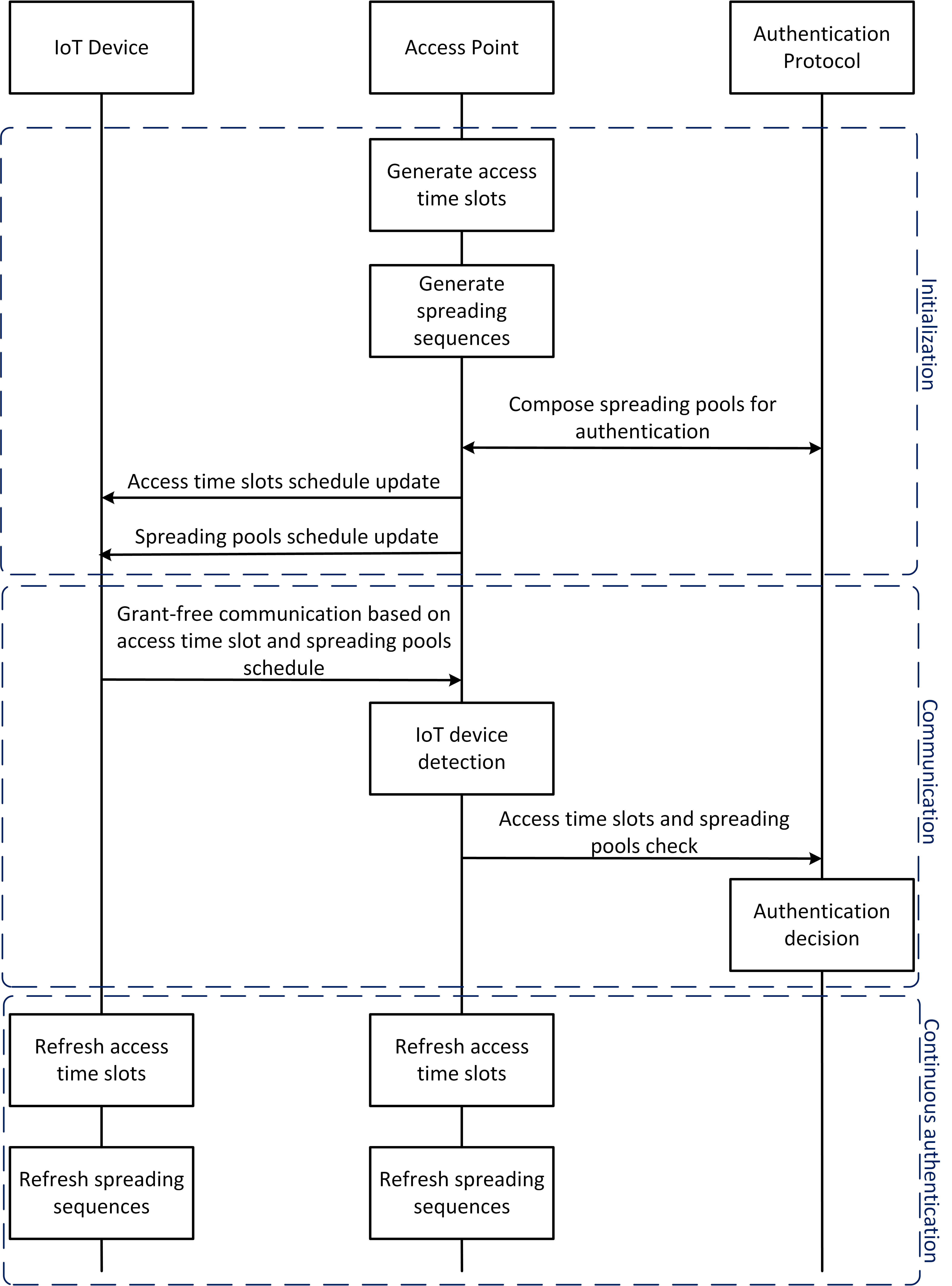} 
\caption{Flowchart of proposed authentication scheme and its interaction with grant-free NOMA system model considered in this work.}
\label{framework}
\end{figure*}

The proposed authentication scheme consists of four processes: (A) access time slots generation, (B) spreading pool construction, (C) seed generation, and (D) authentication decision, and is summarised in Fig.~\ref{PRBS}. The four processes form a cohesive, secure, continuous authentication system between an AP and IoT devices. The AP shares initial access time slots and a codebook matrix with the IoT devices in the initialization stage. The AP and IoT devices then independently refresh the access time slots and generate spreading pools from the codebook matrix. These spreading pools generate the seed for the next round of access time slots. When the current access time slots are exhausted, the seed is refreshed independently at the AP and IoT devices. Finally, the AP uses the access time slots and spreading sequences to make authentication decisions without needing optimal threshold value updates. By combining these processes, the system ensures that the AP and IoT devices can securely communicate, authenticate each other, and maintain continuous authentication over time. This interaction of the proposed authentication scheme with the grant-free NOMA system is illustrated in Fig.~\ref{framework}. The four processes are further explained in detail below.

\subsection{Access Time Slots Generation}
The access time slots for IoT device transmission are divided into recurring time slots of fixed length \cite{9378574}, as depicted in Fig.~\ref{SystemMod}. The IoT devices transmit their signals to the AP in time slots pre-agreed upon between the IoT devices and the AP. Therefore, the AP can quickly identify an illegitimate device based on its time slot access. If the seeds are hidden from illegitimate devices, the access time slots are highly unpredictable. More importantly, a seed can generate several access time slots, allowing each IoT device at the AP to be identified continuously for an extended period. Unlike conventional key-based physical-channel schemes, authentication via access time slots does not entail complex computation or high latencies because, in key-based schemes, access to a coherent key is required for every message transmission. In contrast, the access time slots do not require a shared key for every transmission since the transmission schedules are followed by the IoT device and verified by the AP. Thereby, continuous and lightweight authentication between a transceiver is achieved.

The access time slots are generated using linear feedback shift registers, which entails a statistical behaviour close to truly random sequences and does not entail expensive exponential or modulo operations \cite{bardell1987built}. Therefore, the access time slots are lightweight and challenging to predict by adversaries who do not know the pseudo-random transmission schedules. The access time slots are generated using a monic polynomial of degree $\mu$, which is a prime number with $2^{\mu} - 1$ the maximum length of the generated access time slots \cite{fraleigh2003first}. The generating monic polynomial for a generic variable $\varkappa$ is given as
\begin{equation} \label{monic}
\begin{split}
     f(\varkappa) = C_0 + C_1 \varkappa + C_2 \varkappa^2 + \hdots + C_{\mu} \varkappa^{\mu} =& \sum_{i=1}^{i=\mu} C_i \varkappa^i, \\ & (C_0 = C_{\mu} = 1),
\end{split}
\end{equation}
where $C_0, C_1, \hdots, C_\mu$ are the constant coefficients of the polynomial. It should be noted that it is difficult for illegitimate devices to predict the monic polynomial function used for the access time slot generation since the AP and IoT devices can refresh the monic polynomial, further enhancing the authentication performance. Furthermore, the process of access time slots generation using the monic polynomial in (\ref{monic}) is repeated independently at the AP and $k$-th IoT device to renew the transmission schedule for continuous authentication, provided they have access to an identical seed. Therefore, a transceiver pair does not have to carry out complex hash function operations for seed concealment and sharing, testifying to the low complexity and lightweight nature of the proposed authentication scheme.

\subsection{Spreading Pool Construction}
In this work, we consider that the transmission symbols of the IoT devices are spread with a family of short complex-valued spreading sequences with low cross-correlation values \cite{yuan2016multi, 10178037}, as shown in (\ref{eqmain2}). This allows for loading more IoT devices in a resource block and reducing implementation complexity. Let $\mathbf{C}_{(4,6)}$ represent a complex-valued codebook matrix to support $K = 6$ devices using $N = 4$ resource blocks in an overloaded\footnote{The overloading factor is defined as the ratio of the number of potential IoT devices to the number of available resource blocks in the system, \emph{i.e.}, overloading factor (\%) = $\frac{K}{N} \times 100$.} scenario, given as

\begin{align} \label{spreadmat}
    \mathbf{C}_{(4,6)} = 
\begin{bmatrix}
w_0 & w_4 & w_3 & w_1 & w_6 & w_5 \\
0        & w_2 & w_6 & w_4 & w_5 & w_0\\
w_4 & w_7 & w_0 & w_3 & w_0 & 0 \\
w_3 & w_0 & w_2 & w_4 & w_3 & w_6
\end{bmatrix},
\end{align}
where $w_n$ is the non-zero elements of the codeword. The non-binary and complex-valued spreading sequences in (\ref{spreadmat}) allow for a higher degree of freedom for loading a larger number of IoT devices, thus providing much more flexibility in spreading sequences design, which is reflected by a high overloading factor and demonstrates a true sense of grant-free transmission\footnote{The design of the codebook matrix can be carried out in different ways \cite{shahab2020grant, 6966170, wang2015comparison, 8704942} to enhance the overloading factor of the system further. However, this is outside the scope of this work.}.

Conventionally in grant-free systems, the codebook matrix in (\ref{CodeBookwa}) is stored locally with the AP and shared with all IoT devices independently in the initialization stage, which is later utilized by the IoT devices for data transmission \cite{liu2018sparse, shahab2020grant}. With this sense of practicality in mind, we propose constructing a lightweight mechanism to utilize the codebook matrix in (\ref{CodeBookwa}) for enhanced authentication. This involves constructing spreading pools from the codebook matrix in (\ref{CodeBookwa}) for every IoT device in a cell. Let $\gamma_k$ denote the spreading pool constructed using the codebook matrix $\mathbf{C}_{(4,6)}$ in (\ref{spreadmat}) for the $k$-th IoT device. As such, for the overloaded scenario in (\ref{spreadmat}), the respective spreading pools for $K = 6$  IoT devices can be constructed as
\begin{equation} \label{spreadpool}
    \begin{split}
        \gamma_1 &= \{ w_0, 0, w_4, w_3   \},  \\
        \gamma_2 &= \{ w_4, w_2, w_7, w_0    \},  \\
        \gamma_3 &= \{ w_3, w_6, w_0, w_2    \}, \\ 
        \gamma_4 &= \{ w_1,  w_4,  w_3, w_4    \}, \\
        \gamma_5 &= \{ w_6, w_5, w_0, w_3    \}, \\ 
        \gamma_6 &= \{ w_5,  w_0,  0, w_6    \}.  
    \end{split}
\end{equation}
Once the spreading pools are constructed, the access time slots  are superimposed over the spreading pools for intelligent transmission and enhanced authentication. Thus, the spreading pools in (\ref{spreadpool}) can therefore be rewritten as
\begin{equation} \label{spreadpool_PRBS}
    \begin{split}
        \gamma_1 &= \{ \overbrace{w_0}^1, \overbrace{0}^1, \overbrace{w_4}^0, \overbrace{w_3}^0   \},  \\
        \gamma_2 &= \{ \overbrace{w_4}^1, \overbrace{w_2}^1, \overbrace{w_7}^0, \overbrace{w_0}^1    \},  \\
        \gamma_3 &= \{ \overbrace{w_3}^1, \overbrace{w_6}^0, \overbrace{w_0}^0, \overbrace{w_2}^0    \}, \\ 
        \gamma_4 &= \{ \overbrace{w_1}^1,  \overbrace{w_4}^1,  \overbrace{w_3}^1, \overbrace{w_4}^0    \}, \\
        \gamma_5 &= \{ \overbrace{w_6}^1, \overbrace{w_5}^0, \overbrace{w_0}^0, \overbrace{w_3}^0    \}, \\ 
        \gamma_6 &= \{ \overbrace{w_5}^1,  \overbrace{w_0}^1,  \overbrace{0}^1, \overbrace{w_6}^0    \}.  
    \end{split}
\end{equation}

From (\ref{spreadpool_PRBS}), it can be seen that by jointly utilizing the spreading pools and access time slots, an enhanced security mechanism can be developed, which provides a higher degree of system efficiency (reduction in spreading sequence collision due to intelligent transmission) and security entropy (a two-step mechanism for device authentication). The utilized spreading pools by the respective IoT devices are then used for seed and refreshed access time slot generation. Herein, it should be noted that a longer length of spreading pool and access time slots results in a higher authentication entropy. However, a shorter length results in lower bit-error-rate (BER) performance. This demonstrates a trade-off between authentication and BER performance which can be controlled based on the network requirements.

\subsection{Seed Generation}
Once the spreading pools and their tagged access time slots are exhausted, the AP and IoT devices need to recreate newer spreading pools and access time slots for continuous authentication. In this regard, the $k$-th IoT device can use its current spreading pool to generate a seed value for the newer pools. Let $(c_{1k}, c_{2k}, \hdots, c_{Nk})$ represent the length of the spreading sequences inside a spreading pool $\gamma_k$, and $(l_{1k}, l_{2k}, \hdots, l_{Lk})$ represent the access time slots of the $k$-th IoT device. Then, we generate the seed by taking the XOR of the access time slots and calculating the $\ell_2$ norm of the tagged spreading sequences. This process for an arbitrary spreading pool $\gamma_k$ of the $k$-th IoT device is as follows:\\
Step 1: Take the original spreading pool and its superimposed access time slots
\begin{equation}
        \gamma_k =  \{ \overbrace{w_0}^1, \overbrace{0}^1, \overbrace{w_4}^0, \overbrace{w_3}^0  \}
\end{equation}
Step 2: Take XOR of the access time slots
\begin{equation}
        \gamma_k =  \{ \overbrace{w_0}^0, \overbrace{0}^0, \overbrace{w_4}^1, \overbrace{w_3}^1     \}
\end{equation}
Step 3: Nullify the spreading sequences under $0$'s
\begin{equation}
        \gamma_k =  \{ \overbrace{0}^0, \overbrace{0}^0, \overbrace{w_4}^1, \overbrace{w_3}^1     \}
\end{equation}
Step 4: Take the sum and $\ell_2$ norm of the spreading sequences under $1$'s to obtain preliminary seed
\begin{equation} \label{squaringSeed}
        \Theta = \left|\left|    w_4+w_3 \right|\right|_2
\end{equation}
Step 5: Take the square of the preliminary seed to obtain the final seed
\begin{equation}
        seed = \Theta^2.
\end{equation}
This process is performed independently at the AP and the IoT devices. It should be noted that steps 4 and 5 depend on the resource availability of the IoT devices. That is to say; if the IoT devices are extremely resource-constrained, the preliminary $\Theta$ can be used for access time slots generation since it averts computationally expensive $\mathcal{O}(L^2)$ operation in step 5, as well as results in a shorter key length. However, step 5 provides a longer key length for increased authentication, thereby providing prolonged authentication. The choice of seed in steps 4 and 5 demonstrates a trade-off between the computational performance and security performance of a transceiver pair. Hence, this process should be well-designed to achieve a better trade-off. Furthermore, it should be noted that, unlike the conventional physical-channel-based schemes, the proposed authentication scheme does not rely on channel probing for seed acquisition, seed reconciliation, or authentication. This means that the seed verification phase, which is required in the conventional physical-channel-based authentication schemes due to either imperfect channel probing or quantization errors, is not needed in the proposed authentication scheme, thus paving the way for a practical, lightweight, and independent authentication mechanism in a grant-free NOMA system.

\subsection{Authentication Decision}
The conventional physical-channel-based authentication schemes rely on quantization-aided hypothesis testing as a decision criterion in (\ref{ProbStat}). However, such benchmarks rely on static statistical properties of the physical channel and cannot account for varying attributes of fast-fading physical-channel characteristics, resulting in misdetection. As opposed to this, the proposed authentication scheme does not rely on a quantization-based threshold as an authentication criterion. Instead, the proposed scheme utilizes a two-step authentication decision process, where the AP first matches the access time slots of the transceiver pair and then compares the spreading sequences of the following transmitting schedule. The two-step authentication process enables mitigating misdetection at the AP and averts false alarms. This authentication process is summarised in Algorithm 1, and the main procedure is presented as follows.
\begin{enumerate}
    \item \textit{Line 2:} The sparse transmitted signal vector in the $j$-th time slot is estimated and detected at the AP by the least squares algorithm as \cite{wang2016dynamic}:
    \begin{equation}
        \hat{\mathbf{x}}^{[j]} = \left( \mathbf{G}^{[j]} \right)^\dagger \mathbf{y}^{[j]}.
    \end{equation}

    \item \textit{Line 3:} The codebook matrix $\mathbf{C}^{[j]}$ utilized by the IoT devices in the $j$-th time slot is extracted by applying Hadamard division on the channel matrix as:
    \begin{equation}
        \mathbf{C}^{[j]} = \mathbf{G}^{[j]} \oslash \mathbf{H}^{[j]}.
    \end{equation}

    \item \textit{Line 6:} The spreading pools and the transmission schedule of the $K$ IoT devices is extracted from the codebook matrix in the $j$-th time slot as:
    \begin{equation}
        \gamma_k^{[j](l)}[\text{device}] = \mathbf{C}^{[j]}(:,k).
    \end{equation}

    \item \textit{Line 7-12:} The $l$-th access time slot of the $k$-th IoT device $\gamma_k^{[j](l)} [\text{device}]$ in the $j$-th time slot is compared with the $l$-th access time slot of the AP $\gamma_k^{[j](l)} [\text{AP}]$ in the $j$-th time slot. If the access time slot matches, the authenticated devices indicator function $\mathbf{\Gamma}^{[j]}_k$ for the $k$-th device in the $j$-th time slot is set to $1$. Otherwise, the indicator function records a $0$, deeming the $k$-th device as illegitimate.

    \item \textit{Line 13-17:} The $l$-th spreading sequence of the $k$-th IoT device $\gamma^{[j]}(k,l) [\text{device}]$ from the extracted spreading pool in the $j$-th time slot is compared with the $l$-th spreading sequence of the AP $\gamma^{[j]} (k,l) [\text{AP}]$ in the $j$-th time slot. If the spreading sequence matches, the authenticated devices indicator function $\mathbf{\Gamma}^{[j]}_k$ for the $k$-th device in the $j$-th time slot is set to $1$. Otherwise, the indicator function records a $0$, deeming the $k$-th device as illegitimate.

    \item \textit{Line 20:} The authenticated devices data $\tilde{\mathbf{x}}^{[j]}$ in the $j$-th time slot is determined by calculating the Hadamard product between the estimated sparse transmitted signal vector $\hat{\mathbf{x}}^{[j]}$ and the authenticated devices indicator function $\mathbf{\Gamma}^{[j]}$ in the $j$-th time slot, given as:
    \begin{equation}
        \tilde{\mathbf{x}}^{[j]} = \hat{\mathbf{x}}^{[j]}  \odot \mathbf{\Gamma}^{[j]}.
    \end{equation}
    \end{enumerate}
At the end of the iteration, the authenticated devices data $\tilde{\mathbf{x}}^{[j]}$ in the $j$-th time slot is transformed into a sparse vector, where the data of the illegitimate devices is replaced with $0$'s, whereas the authenticated devices data is recovered. 

\begin{algorithm}[t]
		\caption{The Proposed Authentication Scheme.}
        \textbf{Input:}\\ 
         Received signals: $\mathbf{Y} = \left[\mathbf{y}^{[1]}, \mathbf{y}^{[2]}, \hdots, \mathbf{y}^{[J]} \right] $;\\
         Equivalent channel matrices: $\bar{\mathbf{G}} = \left[\mathbf{G}^{[1]}, \mathbf{G}^{[2]}, \hdots, \mathbf{G}^{[J]}\right]$. 
		
		\textbf{Output:}\\ 
		Authenticated devices indicator: $\bar{\mathbf{\Gamma}} = \left[ \mathbf{\Gamma}^{[1]}, \mathbf{\Gamma}^{[2]}, \hdots, \mathbf{\Gamma}^{[J]} \right] $;\\
		Authenticated devices symbols: $\tilde{\mathbf{X}} = \left[ \tilde{\mathbf{x}}^{[1]}, \tilde{\mathbf{x}}^{[2]}, \hdots, \tilde{\mathbf{x}}^{[J]} \right] $.      
		
	\begin{algorithmic}[1]
	        \Statex \textbf{\underline{Device detection}}
	        \For{$j = 1$ to $J$} \textbf{do}
	            \State $\hat{\mathbf{x}}^{[j]} = \left( \mathbf{G}^{[j]} \right)^\dagger \mathbf{y}^{[j]}$
	            \State $\mathbf{C}^{[j]} = \mathbf{G}^{[j]} \oslash \mathbf{H}^{[j]}$

            \Statex \textbf{\underline{Device authentication}}
            
            \For{$l = 1$ to $L$} \textbf{do}
            \For{$k = 1$ to $K$} \textbf{do}
            \State $\gamma_k^{[j](l)}[\text{device}] = \mathbf{C}^{[j]}(:,k)$.
            
            \Statex \textit{Step 1: (Access time slot check)}
            \If {$\gamma_k^{[j](l)} [\text{AP}]$  $==$ $\gamma_k^{[j](l)} [\text{device}]$}
			    \State $\mathbf{\Gamma}^{[j](l)}_k = 1$.
			\Else 
			    \State $\mathbf{\Gamma}^{[j](l)}_k = 0$.
			    \State Skip to line 17.
			\EndIf			
            
            \Statex \textit{Step 2: (Spreading sequence check)}
            \If {$\gamma^{[j]} (k,l) [\text{AP}]$  $==$ $\gamma^{[j]}(k,l) [\text{device}]$} 
			    \State $\mathbf{\Gamma}^{[j](l)}_k = 1$.
			\Else 
			    \State $\mathbf{\Gamma}^{[j](l)}_k = 0$.
			\EndIf
			\EndFor
			\EndFor
			\State $\tilde{\mathbf{x}}^{[j]} = \hat{\mathbf{x}}^{[j]}  \odot \mathbf{\Gamma}^{[j]}$.
			\EndFor
			
\end{algorithmic}
\textbf{Return:} \\$\bar{\mathbf{\Gamma}} = \left[ \mathbf{\Gamma}^{[1]}, \mathbf{\Gamma}^{[2]}, \hdots, \mathbf{\Gamma}^{[J]} \right]$;\\
$\tilde{\mathbf{X}} = \left[ \tilde{\mathbf{x}}^{[1]}, \tilde{\mathbf{x}}^{[2]}, \hdots, \tilde{\mathbf{x}}^{[J]} \right]$.  
\end{algorithm}

\section{Security Performance Analysis}
The performance of any new authentication scheme can be assessed using security analysis. A comprehensive formal security analysis often necessitates sophisticated modelling, which entails using advanced mathematical frameworks and cryptographic primitives to replicate potential threat scenarios and evaluate system vulnerabilities. In such modelling, formal methods and symbolic representations are employed to capture and analyze the intricate dynamics of potential attacks and the protective countermeasures of the system. This intricate modelling process aims to uncover hidden vulnerabilities, test the system's resilience against various threats, and derive insights for strengthening the system's defence mechanisms \cite{9023936}. However, a formal security analysis is outside the scope of this work. Instead, similar to \cite{9378574}, the effectiveness of our proposed authentication scheme can be assessed rigorously using performance metrics such as entropy, key space, and computational efficiency. Here is why these metrics are employed:
\begin{itemize}
    \item Entropy: This metric indicates a system's resilience against unauthorized access. Specifically, greater entropy suggests that an illegitimate device would be computationally arduous to predict or deduce the system's state.
    \item Key Space: This metric represents the total set of potential keys that could be employed within the system, offering a quantifiable measure of its complexity against brute-force attacks.
    \item Lightweight: This metric aims to minimize computational demands and resource consumption while maintaining stringent security standards.
\end{itemize}
By focusing on these metrics, we can demonstrate the robustness and security performance of the proposed authentication scheme.

\subsection{Entropy} 
Legitimate IoT devices go through periodic updates of the access time slots and spreading pools; therefore, it is challenging for illegitimate devices to spoof the AP. Furthermore, since a transceiver pair independently but identically utilizes multiple spreading sequences from the spreading pool for seed generation, they are difficult for illegitimate devices to predict. Following this, it is clear that the seed is concealed from an adversary if it does not know the access time slots and the corresponding spreading pools. Furthermore, updating the access time slots and spreading pools will provide further protection for legitimate IoT devices by renewing their access sequences over time. Hence, the proposed authentication scheme provides enhanced protection against spoofing attacks and pertains to legitimate communications between IoT devices and the AP.

With this understanding, entropy is defined as a metric that measures the uncertainty associated with the randomness of a system \cite{wang2015survey} and is used to evaluate the security strength of the authentication scheme. Thus, entropy is defined as
\begin{equation} \label{entro1}
    E_{total} = \sum_{r=1}^R E_r,
\end{equation}
where
\begin{equation}\label{entro2}
    E_r = -p_{r0} \log p_{r0} - (1 - p_{r0}) \log (1-p_{r0}).
\end{equation}
$R$ represents the total length of the shared key, and $p_{r0}$ denotes the posterior probability of the $r$-th bit when it is $0$ from the illegitimate devices’ knowledge.

\begin{lemma}
\textit{The entropy of the proposed authentication scheme is higher than that of the physical-channel key generation schemes of \cite{wang2015survey, wilhelm2013secure, edman2016security, zhang2016efficient}.}
\end{lemma}

\textit{Proof:} We provide proof using heuristic arguments as follows. Assuming $R$ denotes the length of the access time slots and the key in the physical-channel key generation schemes, we denote $p_{r0}^{\text{I}}$ and $p_{r0}^{\text{II}}$ as their posterior probabilities of the $r$-th bit when it is $0$ from the illegitimate devices’ knowledge, respectively. It should be noted that the proposed authentication scheme relies on multiple attributes, \emph{i.e.}, it utilizes $N$ spreading sequences for seed generation. On the contrary, the physical-channel key generation schemes rely on a single attribute for shared key generation. Since multiple attributes are being utilized in the proposed authentication scheme and legitimate IoT devices follow the pre-agreed access time slots for transmission, it is difficult for illegitimate devices to spoof the AP. Then, 
\begin{equation} \label{Entro1G2}
    \left| p_{r0}^{\text{I}} - \frac{1}{2}    \right|   <    \left|p_{r0}^{\text{II}}  - \frac{1}{2}    \right|
\end{equation}
holds \cite{9378574}, which means the illegitimate devices have less knowledge that the $r$-th bit is $0$ in the proposed authentication scheme. Let $E_r^{\text{I}}$ denote the entropy of the proposed authentication scheme, and $E_r^{\text{II}}$ denote the entropy of the physical-channel key generation schemes. Then, from (\ref{Entro1G2}), we can concur that
\begin{equation}\label{entroI}
    E_r^{\text{I}} > E_r^{\text{II}}
\end{equation}
holds. This completes the proof.

\subsection{Key Space}
Due to their limited computational resources, the resource-constrained IoT devices cannot compute shared keys for every data transmission, required by conventional encryption methods. To overcome this inherent issue, resource-constrained IoT devices rely on shortened keys to reduce the computational overhead. However, shortened keys can be more vulnerable to malicious attacks as they can be easily cracked by attackers using brute force. This is because sophisticated attackers with rapidly growing processing power can compromise the short-length keys within a much shorter time than before, for example, by using exhaustive search approaches \cite{zeng2015physical}. Therefore, an additional layer of security based on low computational cost is required. Based on multi-factor attributes, the proposed authentication method complements the overall security paradigm by acting as another source of randomness to provide additional entropy to the system. This authentication at the lower layer compensates for entropy loss due to the use of shortened keys in the higher layers in resource-constrained IoT devices.

\begin{lemma}
\textit{The key space of the proposed authentication scheme is higher than that of the physical-channel key generation schemes of \cite{wang2015survey, wilhelm2013secure, edman2016security, zhang2016efficient}.}
\end{lemma}

\textit{Proof:} We provide proof using heuristic arguments as follows. Assuming that $R$ represents the length of the key in the proposed authentication scheme and physical-channel key generation schemes, we denote $\kappa_R^{\text{I}}$ and $\kappa_R^{\text{II}}$ as the upper bound of the key search space, respectively. We know that the proposed authentication scheme utilizes the access time slots and the complex spreading sequences for IoT device authentication. On the other hand, the physical-channel key generation schemes rely on the attribute of the physical channel for key generation. Thus, in Table \ref{TabKeySpace}, we demonstrate the key search space versus the key length of the proposed authentication scheme and physical-channel key generation schemes. It is evident that the proposed authentication scheme achieves a higher search space than the physical-channel key generation schemes for the same key length. This is because the proposed technique utilizes complex spreading sequences and access time slots, which adds another source of randomness to the system for key generation. Therefore, the proposed authentication scheme is less susceptible to brute force attacks than the physical-channel key generation schemes for the same key length. Thus, from Table \ref{TabKeySpace}, it is concurred that
\begin{equation} \label{entroII}
    \kappa_R^{\text{I}} > \kappa_R^{\text{II}}
\end{equation}
holds. This completes the proof. 

Since the proposed authentication scheme introduces more randomness into the network, the total system entropy $E_{total}$ is higher than physical-channel key generation schemes. Hence, the proposed authentication scheme can be integrated into the network to provide additional entropy for improving the system's resistance to attacks.

\definecolor{MineShaft}{rgb}{0.2,0.2,0.2}
\begin{table}
\centering
\caption{Key length versus search space complexity of physical-channel-based and proposed techniques.}
\label{TabKeySpace}
\resizebox{\linewidth}{!}{%
\begin{tblr}{
  cell{1}{1} = {r=2}{},
  cell{1}{2} = {c=2}{},
  cell{1}{4} = {c=2}{},
  cell{2}{3} = {fg=MineShaft},
  cell{2}{5} = {fg=MineShaft},
  cell{3}{3} = {r=5}{c},
  cell{3}{5} = {r=5}{c},
  vlines,
  hline{1,3,8} = {-}{},
  hline{2} = {2-5}{},
  hline{4-7} = {1-2,4}{},
}
Key length & Physical-channel key generation schemes &                                                         & Proposed authentication scheme &                                                         \\
           & Search space                            & Authentication~complexity & Search space                   & Authentication~complexity \\
9          & 512                                     & $\mathcal{O}(\mathcal{N})$                              & 8192                           & $\mathcal{O}(1)$                                        \\
11         & 2048                                    &                                                         & 32768                          &                                                         \\
13         & 8192                                    &                                                         & 131072                         &                                                         \\
15         & 32768                                   &                                                         & 524288                         &                                                         \\
17         & 131072                                  &                                                         & 2097152                        &                                                         
\end{tblr}
}
\end{table}

\subsection{Lightweight}
The proposed authentication scheme utilizes the transmission parameters and access time slots for IoT device authentication. Conversely, the proposed authentication scheme does not rely on physical-channel probing for IoT device authentication. As a result, the seed verification phase is not required in our proposed authentication scheme. More importantly, the proposed schemes provide continuous authentication by checking the spreading sequences and access time slots of the IoT devices instead of generating and verifying shared keys repeatedly. As a result, as shown in Table~\ref{TabKeySpace}, compared to the physical-channel-based key generation schemes, the proposed authentication schemes achieve a lower authentication complexity for $\mathcal{N}$ times of authentication, which validates the lightweight nature of the proposed authentication scheme.

\section{Results and Discussion}
In this section, we evaluate the performance of the proposed authentication scheme in solving the device authentication problem. We plot the performance of three physical-channel-based authentication benchmark solutions: using binary hypothesis testing (BHT) \cite{xie2021physical}, using machine learning-based SVM \cite{9354771}, and using deep neural network-based (NN) detection \cite{8935162}. For these three benchmark solutions, the core architectures are borrowed from the respective works but their input configurations have been adjusted to our system model for a fair comparison. For these benchmark solutions, the estimates of the received signal strength indicator (RSSI), the channel impulse response (CIR), and the channel frequency response (CFR) are used as attributes from the physical channel for authentication \cite{9279294}. Specifically, due to the correlation of adjacent CIRs and CFRs on the same path, the temporal process of the $i$-th subpath at the $j$-th time slot is given as \cite{9279294}
\begin{equation}
    h_{i}(j)=\zeta h_{i}(j-1)+\sqrt{(1-\zeta^2) \sigma_{i}^{2}} u_{i}(j-1),
\end{equation}
where $\zeta \in [0,1]$ represents the physical-channel correlation of two successive subpaths and $u_{i}$ is a driving noise which is modeled as a zero-mean complex Gaussian random variable with unit variance \cite{xie2021physical}. The path loss between the AP and the $k$-th IoT device is modeled as $128.1 + 37.6 \log_{10} (d_i)$, where $d_i$ is the distance (in km) \cite{ETUR2021evolved}. Additionally, for the benchmark schemes, the physical channels of the illegitimate devices are assumed to be independent of the legitimate IoT devices, meaning the illegitimate devices are assumed to be at a distance greater than half wavelength from the legitimate IoT devices.

Assuming initial authentication between a transceiver pair in the $j$-th time slot, their observation characteristics are shown in Table \ref{tab:SumRes}. As detailed in section III-C, the AP and IoT device independently extract the seed by utilizing the spreading pool used for data transmission. Since the seed source is the spreading pool, extracted from the codebook matrix and available with the transceiver pair locally, there is no requirement for seed verification. Therefore, once the seed is acquired, the AP and IoT independently generate the access time slots required for transmission. In this work, we utilize the following monic polynomial for the access time slots generation
\begin{equation}
    f(\varkappa) = 1 + \varkappa^1 + \varkappa^3. 
\end{equation}

\begin{table}[t]
\centering
\caption{Access time slots generation using seed.}
\label{tab:SumRes}
\resizebox{\linewidth}{!}{%
\begin{tabular}{|l||l|} 
\hline
State                                                                                        & Observations                                                                                                                                                                                                                                                                                               \\ 
\hline
\begin{tabular}[c]{@{}l@{}}Spreading pool utilized \\between a transceiver pair\end{tabular} & $\gamma = \{ -4-4i, -0+8i, 1-1i, \hdots, -2+2i, 4 \}$                                                                                                                                                                                                                                                      \\ 
\hline
Seed extracted by the AP                                                                     & $1 1 1 1 0 1 0 0 0 0 0 0 0$                                                                                                                                                                                                                                                                                \\ 
\hline
Seed extracted by the IoT                                                                    & $1 1 1 1 0 1 0 0 0 0 0 0 0$                                                                                                                                                                                                                                                                                \\ 
\hline
\begin{tabular}[c]{@{}l@{}}Access time slots at the AP \\and IoT\end{tabular}                & \begin{tabular}[c]{@{}l@{}}$1 0 0 0 0 1 0 0 0 0 0 0 1 1 1 0 1 1 0 1 0 0 1 0 1 0 1 0 1 1 0 0 1 1$ \\$1 1 0 0 0 1 1 1 0 0 0 0 1 0 0 1 0 0 1 1 1 1 0 1 1 1 0 1 0 0 1 1 1 0$\\$1 1 1 1 1 0 0 0 1 1 1 0 0 1 0 1 1 1 1 0 0 1 1 1 1 1 1 0 0 1 0 0 1 1$\\$1 1 0 1 1 0 0 1 1 1 1 0 0 1 0 0 0 0 0 0 1$\end{tabular}  \\
\hline
\end{tabular}
}
\end{table}

\subsection{Experimental Setup}
In the simulations, unless otherwise stated, $K = 200$ potential devices simultaneously share $N = 100$ resources. Thus, the overloading factor (OF) is 200\%. For every time slot, there is $S = 20$ number of active devices randomly selected from the set $\{1, 2, \hdots, K\}$. The number of time slots is fixed at $J = 7$. The transmitted signals are modulated by Quadrature Phase Shift Keying. The signal-to-noise ratio (SNR) range is set between 0 to 25 dB. The oracle least squares algorithm is utilized for device detection.

The simulations are carried out on the Gadi supercomputer of the National Computational Infrastructure (NCI), Australia, utilizing 48 cores of Intel Xeon Platinum 8274 (Cascade Lake) processors and 192GB of random access memory. The simulations are carried out on MATLAB 2021b. The results are averaged over 1000 Monte Carlo trials.

\begin{figure}[t] 
\centering
\includegraphics[scale=0.6]{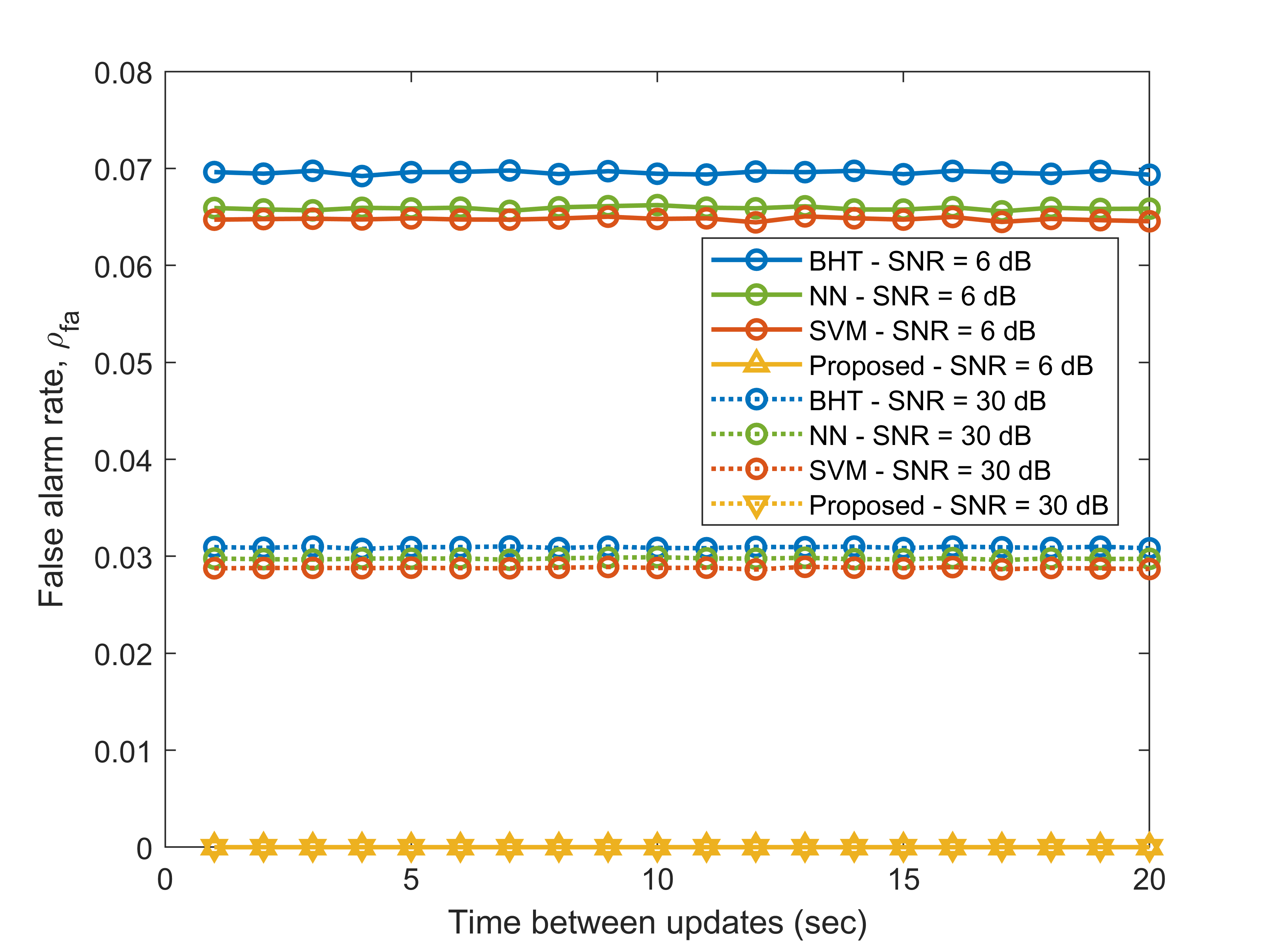}
\caption{False alarm rate, $\rho_{\mathrm{fa}}$, versus the time between updates (sec), with the total number of potential devices $K = 200$, the number of resources $N = 100$, and the number of active devices $S = 20$.}
\label{falsealarmrate}
\end{figure}

\subsection{Performance Metrics}
In order to appropriately evaluate the authentication performance, we use the following metrics: the false alarm rate ($\rho_{\mathrm{fa}}$), the misdetection rate ($\rho_{\mathrm{md}}$), and the spreading sequence collision rate ($\rho_{\mathrm{sc}}$) as performance metrics. Given the transmit signal $\mathbf{x}$, authenticated devices data $\tilde{\mathbf{x}}$, the authenticated devices indicator $\mathbf{\Gamma}$, and the spreading pool $\mathbf{\gamma}$ for the $k$-th IoT device in the $j$-th time slot, the performance metrics are defined as follows.
\begin{itemize}
    \item False alarm rate: This metric evaluates the rate of legitimate IoT devices being falsely detected as illegitimate devices, given as
    \begin{equation}
        \rho_{\mathrm{fa}} =  \frac{1}{K} \displaystyle\sum_{k \in \mathbf{x}^{[j]}} P \left\{  \mathbf{\Gamma}_k^{[j]} = 0 \; | \; \mathbf{x}_k^{[j]} = 1   \right\}.
    \end{equation}
    
    \item Misdetection rate: This metric evaluates the rate of illegitimate IoT devices being misdetected, given as
    \begin{equation}
        \rho_{\mathrm{md}} =  \frac{1}{K} \displaystyle\sum_{k \in \mathbf{x}^{[j]}} P \left\{  \mathbf{\Gamma}_k^{[j]} = 1 \; | \; \mathbf{x}_k^{[j]} = 0   \right\}.
    \end{equation}
    
    \item Spreading sequence collision rate: This metric evaluates the rate of legitimate IoT devices utilizing the same spreading sequence in the same access time slot, given as:
    \begin{equation}
        \rho_{\mathrm{sc}} =  \frac{1}{K} \displaystyle\sum_{k \in \tilde{\mathbf{x}}^{[j]} } P \left\{  \gamma_k^{[j](l)} == \gamma_{i \neq k}^{[j](l)}  \right\}.
    \end{equation}
\end{itemize}

\subsection{Authentication Performance}
Fig.~\ref{falsealarmrate} plots the false alarm rate, $\rho_{\mathrm{fa}}$, versus the time between updates (sec) for $K = 200$, $N = 100$, and $S = 20$. The false alarm events are avoided in the proposed authentication scheme due to the spreading sequences-based seed generation technique proposed in this paper. The spreading sequences-based seed generation allows AP and IoT devices to independently acquire identical seeds for the access time slots generation. In essence, the access time slots generated in the proposed authentication scheme between the AP and an IoT device are identical and do not require parity bits for seed reconciliation. On the contrary, since the benchmark schemes rely on estimates of multiple attributes of the physical channel, false alarm events are inevitable due to the imperfect and time-varying nature of the physical channel encountered due to reliance on the randomness of the channel for seed acquisition. Moreover, lower SNR could lead to a higher false alarm rate in physical-channel-based schemes since its performance explicitly relies on observing physical-channel attributes.

\begin{figure}[t] 
\centering
\includegraphics[scale=0.6]{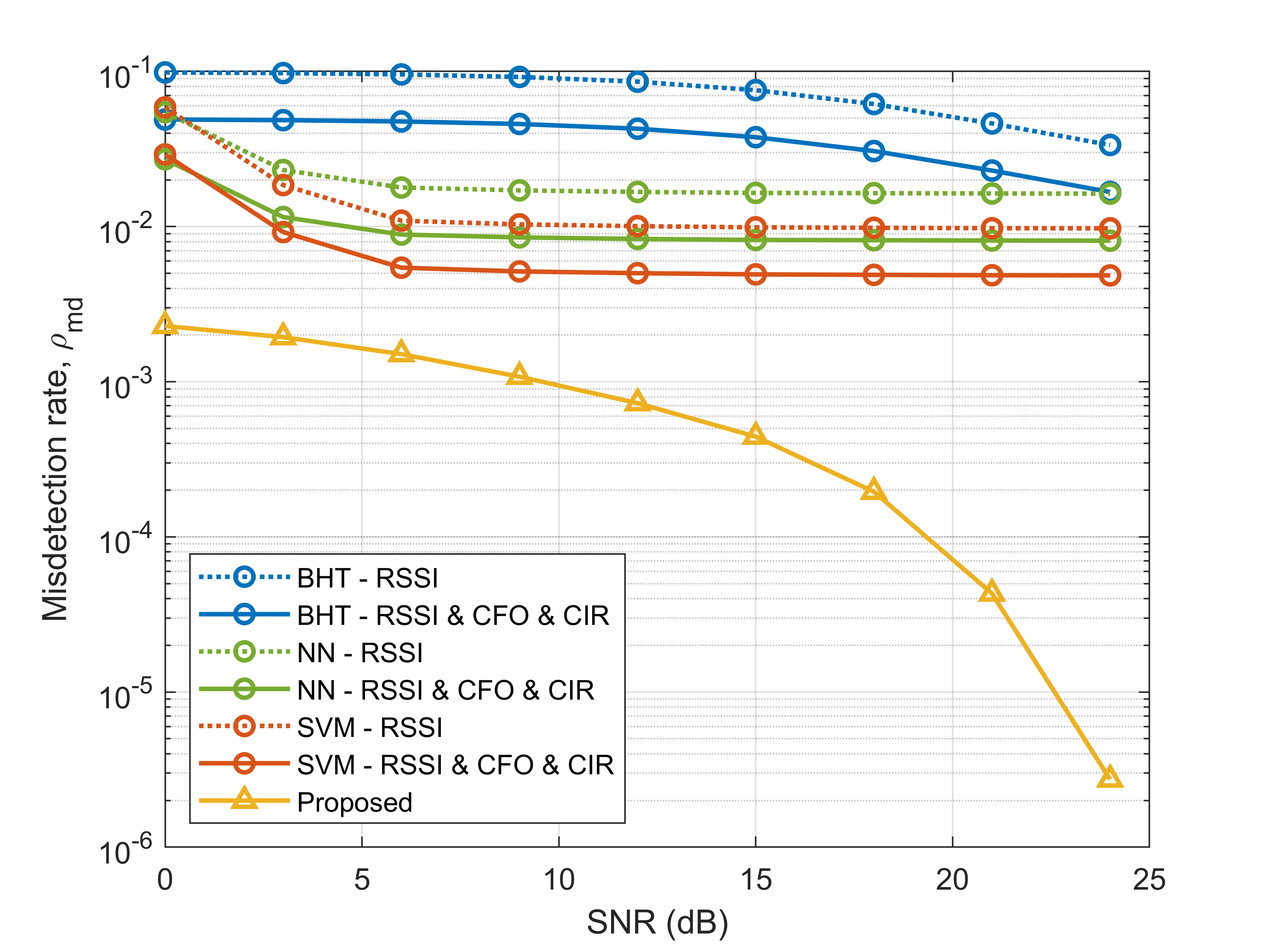}
\caption{Misdetection rate, $\rho_{\mathrm{md}}$, versus SNR (dB), with the total number of potential devices $K = 200$, the number of resources $N = 100$, and the number of active devices $S = 20$.}
\label{misdetectionrate_notemp}
\end{figure}

Fig.~\ref{misdetectionrate_notemp} plots the misdetection rate, $\rho_{\mathrm{md}}$, versus SNR (dB) for $K = 200$, $N = 100$, and $S = 20$\footnote{Fig.~\ref{misdetectionrate_notemp} is simulated with 100,000 Monte Carlo trials to evaluate its performance for the entire SNR range. This simulation took 19 hours to execute on the Gadi NCI supercomputer.}. We can observe that in the entire SNR range, the proposed authentication scheme's misdetection rate decreases and achieves a near-threefold performance gain against the benchmark schemes at the higher SNR range. For instance, the performance gain is around 10 dB compared to the traditional BHT-based authentication scheme at SNR = 6 dB. This trend is because the AP and IoT devices identically but independently generate the access time slots using the spreading sequences. These spreading sequences and the access time slots are then used for IoT device authentication. Hence, the proposed authentication scheme is robust in the noisy wireless communication environment. Fig.~\ref{misdetectionrate_notemp} also demonstrates the authentication performance of the benchmark schemes for single and multiple attributes, which rely on estimates of these attributes from the physical channel for device authentication. It can be seen that the benchmark schemes have a higher misdetection rate at lower SNR, which is due to the imperfect physical-channel mismatch between the AP and IoT devices, which requires the continuous updating of the decision boundary. More importantly, the reliance of the proposed authentication scheme on spreading sequences for continuous authentication adds an additional element to the authentication mechanism and generally makes it more difficult for an illegitimate device to spoof the AP under the proposed authentication protocol. By employing our proposed authentication scheme, the AP gains the ability to differentiate between legitimate and illegitimate devices based on their utilization of spreading sequences and transmission characteristics. Consequently, our proposed authentication scheme eliminates the disparities introduced by distance-related factors when distinguishing between legitimate and illegitimate devices, and therefore, the correlated physical channel characteristics do not play a role in spoofing the AP.

\subsection{Robustness in Different Configurations}

\begin{figure}[t] 
\centering
\includegraphics[scale=0.6]{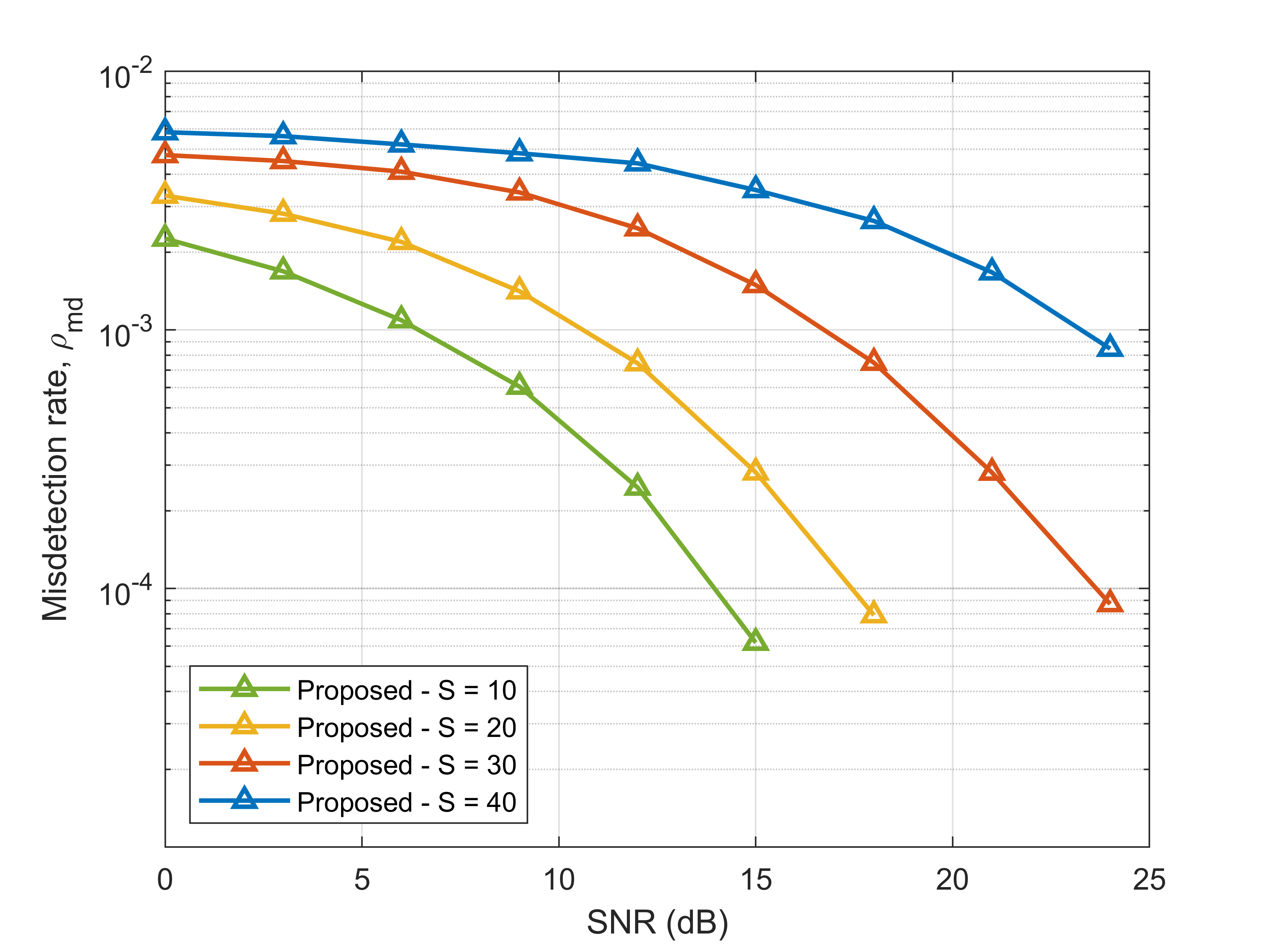}
\caption{Misdetection rate, $\rho_{\mathrm{md}}$, versus SNR (dB) for the varying number of active devices $S$, with the total number of potential devices $K = 200$, and the number of resources $N = 100$.}
\label{misdetectionrate_varyingS}
\end{figure}

Fig.~\ref{misdetectionrate_varyingS} plots the misdetection rate, $\rho_{\mathrm{md}}$, versus SNR (dB) for the varying number of active devices $S$, with $K = 200$, and $N = 100$. It can be seen that the proposed authentication scheme is capable of handling a variety of active transmitting devices $S$. This is because the proposed authentication scheme does not rely on physical channels for binary testing as a decision boundary, which requires an update to the decision boundary for every change in the number of active devices $S$. Since the proposed authentication scheme relies on the spreading sequences extracted from the codebook matrix, the proposed authentication scheme can adapt to any number of active transmitting devices $S$. It should be noted that the reduction in misdetection rate $\rho_{\mathrm{md}}$, caused by the increase in the number of active transmitting devices $S$ is due to the device estimation errors, which is a side effect of the grant-free NOMA system.

\begin{figure}[t] 
\centering
\includegraphics[scale=0.6]{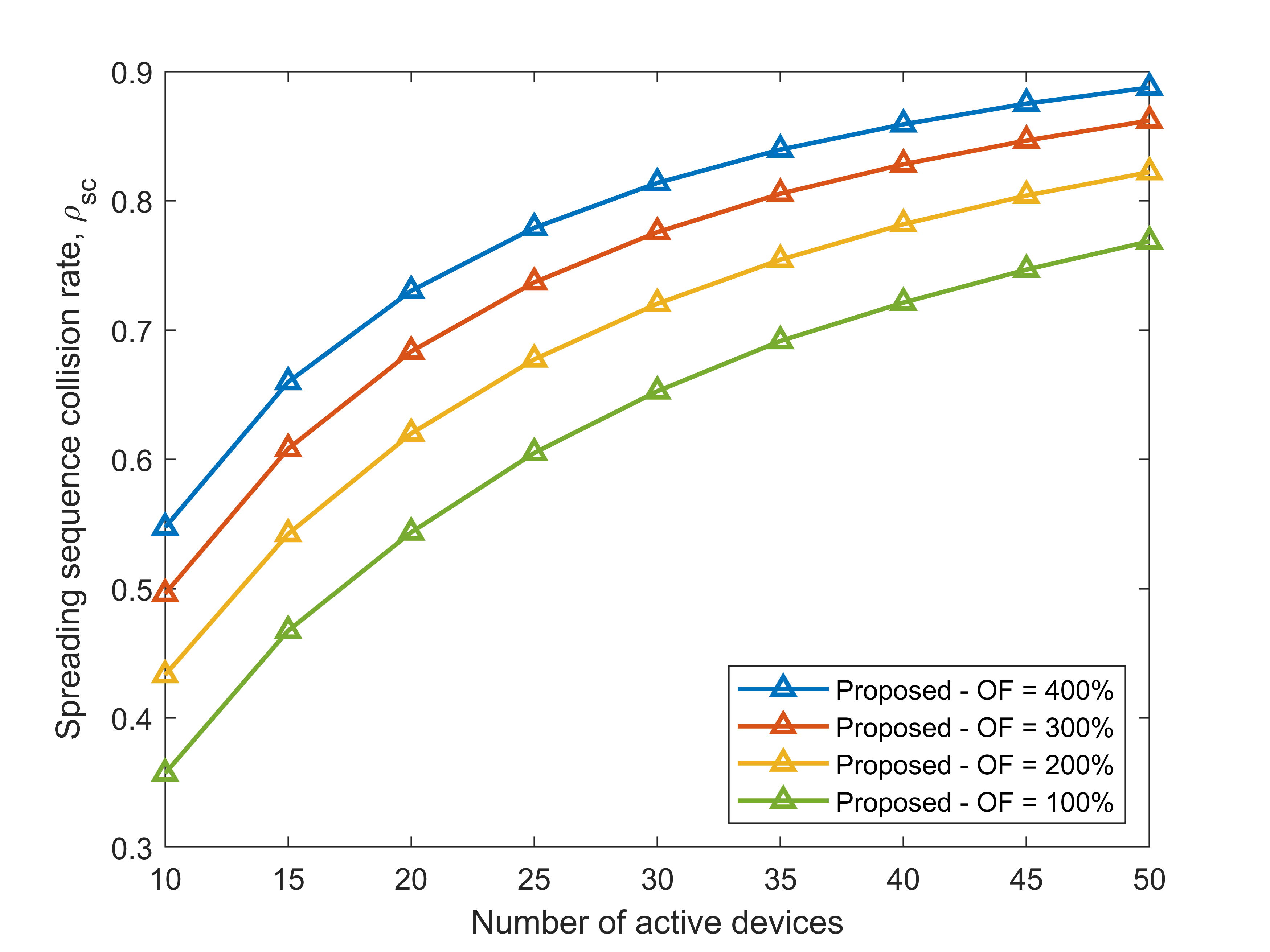}
\caption{Spreading sequence collision rate, $\rho_{\mathrm{sc}}$, versus the varying number of active devices $S$, with the total number of potential devices $K = 200$.}
\label{misdetectionrate_varyingOF}
\end{figure}

Fig.~\ref{misdetectionrate_varyingOF} plots the spreading sequence collision rate, $\rho_{\mathrm{sc}}$, versus the number of active devices $S$ for different OF settings, with $K = 200$. The spreading sequence collision rate $\rho_{\mathrm{sc}}$ increases with the number of active devices $S$. It is also evident that a low number of resources $N$ results in a higher OF, which also increases the spreading sequence collision rate $\rho_{\mathrm{sc}}$. This is because when more active devices $S$ transmit simultaneously with shared resources $N$, the probability of the two or more active devices using the same resource for transmission increases, which increases the spreading sequence collision rate $\rho_{\mathrm{sc}}$. It should be noted that these collisions result from the system's bottleneck due to the inherent nature of the grant-free NOMA systems. Even so, the proposed authentication scheme can handle various active devices $S$ and therefore is robust to different system settings.

\begin{figure}[t] 
\centering
\includegraphics[scale=0.6]{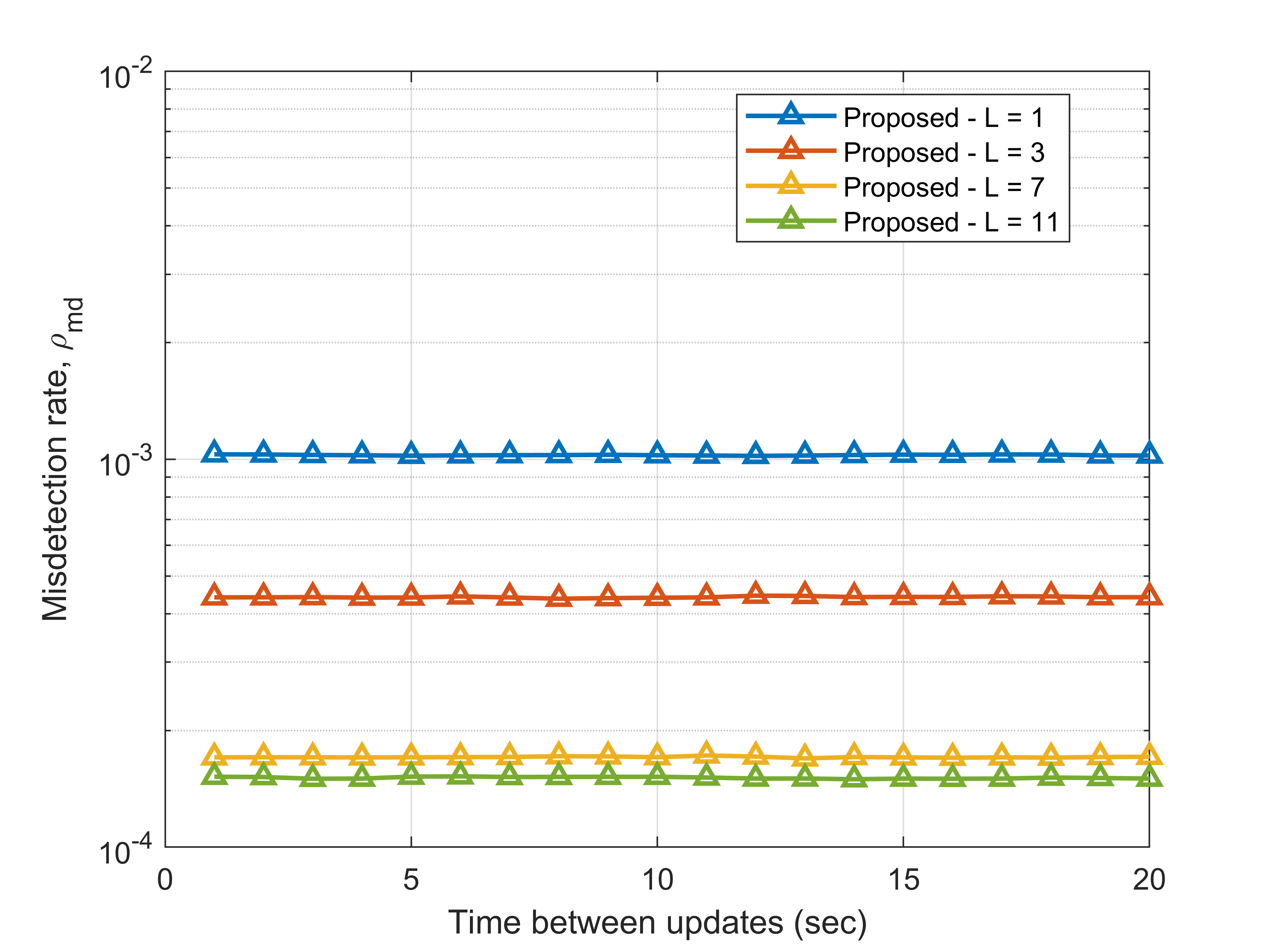}
\caption{Misdetection rate, $\rho_{\mathrm{md}}$, versus the time between updates (sec) for the varying length of authentication sequence $L$, with the total number of potential devices $K = 200$, the number of resources $N = 100$, and the number of active devices $S = 20$.}
\label{misdetectionrate_varyingOF_2}
\end{figure}

Fig.~\ref{misdetectionrate_varyingOF_2} plots the misdetection rate, $\rho_{\mathrm{md}}$, versus the time between updates (sec) for $K = 200$, $N = 100$, and $S = 20$. It can be seen that with the increase in the length $L$ of the access time slots, the misdetection rate $\rho_{\mathrm{md}}$ of the proposed authentication scheme decreases. This is because the longer length of access time slots results in a more randomized transmission pattern for legitimate IoT devices, which is difficult for an illegitimate device to predict and spoof the AP. However, shorter lengths of access time slots, which result in a higher misdetection rate, are less computationally expensive to generate. Therefore, the choice between the length of the access time slots and the system's computational requirements is a trade-off that can be carefully chosen, depending on the requirement of the network.

\begin{figure}[t] 
\centering
\includegraphics[scale=0.6]{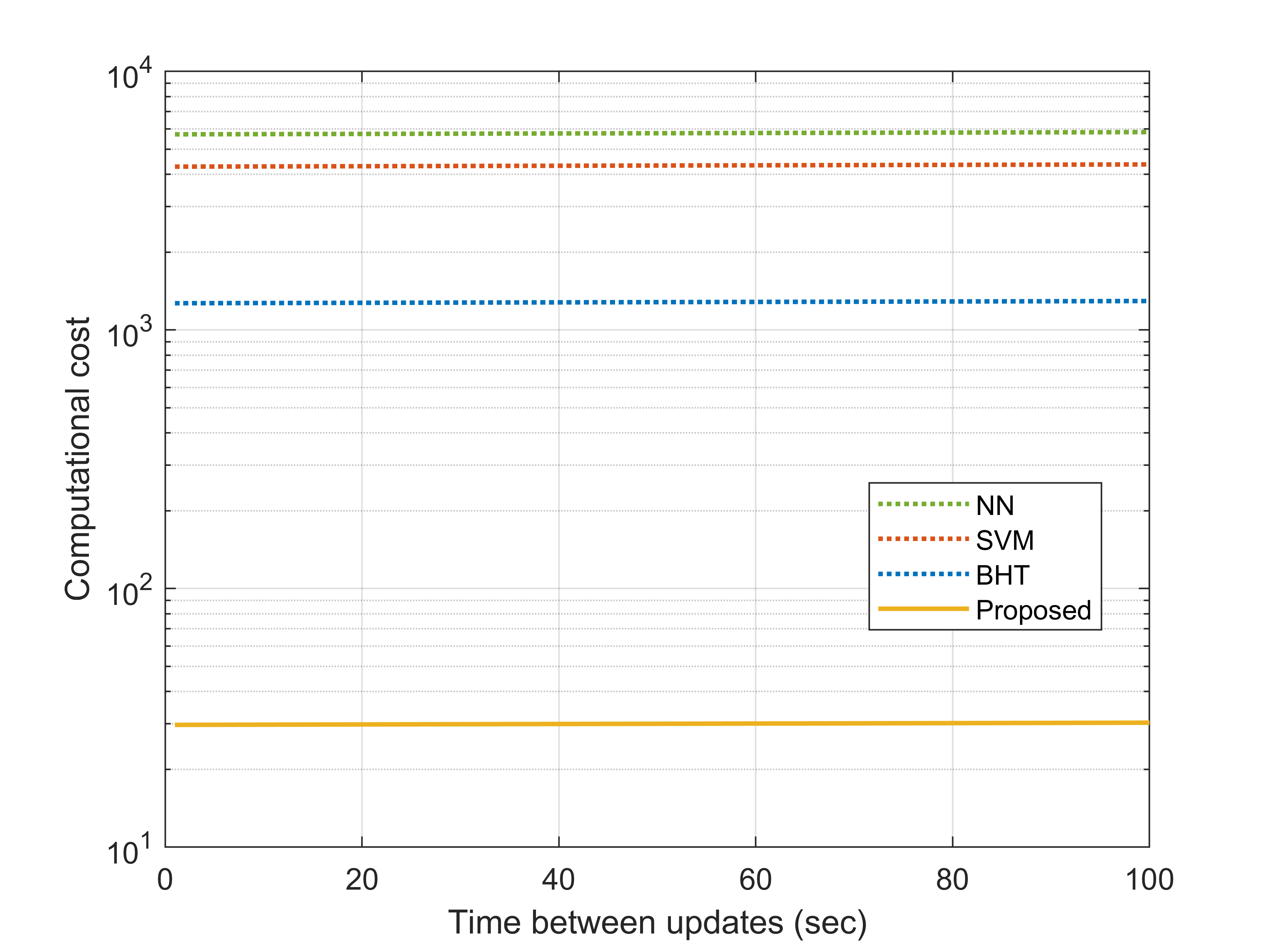}
\caption{Computational cost versus the time between updates (sec), with the total number of potential devices $K = 200$, the number of resources $N = 100$, and the number of active devices $S = 20$.}
\label{ComputeCost}
\end{figure}

Fig.~\ref{ComputeCost} plots the computational cost versus the time between updates (sec) for $K = 200$, $N = 100$, and $S = 20$. It can be seen that the proposed authentication scheme attains a lower computational cost than the benchmark schemes. This is because the proposed authentication scheme relies on the access time slots and the spreading pools as its source of IoT device authentication. Since the codebook matrix, which is utilized to derive the spreading pools, is managed by the AP and does not require creating any threshold boundaries, the proposed scheme has a lower computational cost. On the contrary, the physical-channel-based benchmark schemes rely on a computationally expensive exhaustive search to derive decision boundaries for IoT device authentication. Furthermore, methods such as SVM and hypothesis testing are required for continuous parameter updates due to the time-varying nature of the physical channel for device authentication.

\section{Conclusion and Future Work}
In this paper, we proposed a secure and efficient continuous authentication scheme for IoT devices. Our scheme utilized the grant-free NOMA protocol's transmission characteristics as a source for seed generation and device authentication. By utilizing pre-arranged access time slots and spreading sequences of IoT devices at the AP, the proposed scheme eliminated the need for channel probing, seed reconciliation, and authentication. Simulation results demonstrated the effectiveness of the proposed scheme, with a near three-fold reduction in misdetection rate and close to zero false alarm rate in various system configurations. Additionally, our proposed scheme offered computational efficiency compared to benchmark schemes based on support vector machine and binary hypothesis testing utilizing physical channel information, with at least half the computational cost. 

Future research should explore the extended application of the proposed authentication scheme beyond its current context in signature-based grant-free NOMA schemes, examining its adaptability in various scenarios to understand its effectiveness in diverse wireless communication environments. Additionally, investigating the authentication scheme's implementation in satellite-IoT networks presents an exciting opportunity to address unique challenges related to vast coverage and long-distance communication, potentially unlocking secure and efficient communication in satellite-based IoT applications. To ensure real-world viability, a comprehensive security analysis is crucial, covering a wide range of potential attacks, including adversarial and resource exhaustion attacks, to identify weaknesses and develop robust authentication solutions for IoT devices. Additionally, a formal security analysis of the authentication scheme can be carried out to further understand its workability in different scenarios. Furthermore, scalability should be investigated to ensure efficient authentication, even in massive-scale deployments. By optimizing the scheme without compromising security and addressing these research areas, the groundwork can be laid for secure, adaptive authentication solutions that bolster IoT device security and seamless integration into our interconnected world. 

\balance
\bibliographystyle{IEEEtran}
\bibliography{IEEEabrv, references}

\end{document}